\documentclass[12pt,a4paper,aps,prb,superscriptaddress]{revtex4}

\usepackage{epsfig}

\begin{document}


\thispagestyle{empty}

\title{Alternative Method for Studying Color Centers in CsCl Thin Films} 
\author{Kuldeep Kumar, \& P.Arun}
\affiliation{Department of Physics \& Electronics, S.G.T.B. Khalsa College, 
University of Delhi, Delhi-110 007, India}
 \email{arunp92@physics.du.ac.in}
\author{R.Kumar}
\affiliation{Department of Physics. Ch. Charan Singh University, 
Meerut, Uttar Pradesh, India}
\author{N.C.Mehra}
\affiliation{University Science \& Instrumentation Center, 
University of Delhi, Delhi-110 007, India}
\author{L.Makinistian}
\affiliation{Facultad de Ingenier\'{\i}a,
Universidad Nacional de Entre R\'{\i}os, 
3101 Oro Verde (ER), Argentina}
\affiliation{Grupo de Materiales Computacionales,
INTEC-CONICET,
Guemes 3450- 3000 Sante Fe,
Argentina}
\author{E.A.Albanesi}
\affiliation{Facultad de Ingenier\'{\i}a,
Universidad Nacional de Entre R\'{\i}os, 
3101 Oro Verde (ER), Argentina} 
\affiliation{Grupo de Materiales Computacionales,
INTEC-CONICET,
Guemes 3450- 3000 Sante Fe,
Argentina}

\begin{abstract}
The optical properties of cesium chloride (CsCl) are changed with presence
of vacancies in the crystal structure giving rise to what is called as
"color centers". We have unconventionally adopted Tauc's method to determine 
the characteristic color centers and have modeled the correlation among the 
film optical properties and lattice size, highlighting the interrelation 
between the structural and optical properties of alkali halide films.
\end{abstract}

\maketitle


\section{Introduction}

Excess alkali atoms or vacancies due to lack of halogen atoms in Alkali 
Halides (AH) compounds are known to give color in otherwise transparent
crystals. In the literature such "color centers" are manufactured by X-ray
irradiating AH crystals.\cite{rab, ava} This results in characterisitic peaks with
large absorption in the UV-visible spectra. Usual characterization of color
centers are done by analysing the peak positions whose intensity ofcourse
depends on the number of vacancies manufactured. In this present article we
report our study of color centers in CsCl thin films. The color-centers are
formed due to the thermal evaporation method used in fabricating these thin
films, where some AH molecules might have dissociated on heating. However, 
since the number would have been small, the vacancies developed in turn
would be minimum resulting in the absorption spectra having no distinct 
peaks. We have used the slope of the gradually increasing absorption with
decreasing wavelength to comment on the nature of color centers. This method
though is routinely used to compute the optical band edge, has not be used
in studying color centers.

To relate our experimental findings, we have then theoretically modeled the 
CsCl system. For our calculations we have used the FP-LAPW method within the 
density functional theory (DFT), in the form implemented by the WIEN2k code,
\cite{WIEN2k} a very accurate first-principles scheme 
used in modeling properties of materials. We have used the so called 
generalized gradient approximation (GGA), 
\cite{Perdew-etal92,Perdew-Wang92} in the formal 
parametrization scheme of Perdew-Burke-Ernzerhof (PBE).\cite{PBE96} 
This corrected functional is semi-local and thus more sensitive 
to non-spherical components of the density, resulting in a better 
performance than the local density approximation (LDA) when applied 
in a full potential scheme like the WIEN2k. 
It is to be emphasized, though, that both LDA and GGA always yield
underestimated band gaps. An usual, empirical correction of this is
done by using the scissors operator,\cite{Lanoo84} which basically consists of
adjusting the band gap with a constant potential to reproduce the
experimental energy band gaps. This operator is often used, particularly
in the determination of the band gap offsets\cite{Wei-Zunger97,Albanesi94} 
which appear when considering interfaces between different semiconductors, 
in optical transitions\cite{Laskowski-Christensen05}, 
and also when bulk properties under pressure are studied.\cite{Nabi2000} 
We have applied it, with a constant value of 2.45 eV, to our results of 
densities of states by shifting up the regions above the Fermi energy.

\section{Experimental}
Thin films of Cesium Chloride were grown by thermal evaporation using
99.98\% pure starting material obtained from Loba Chemie Pvt Ltd, Mumbai. 
The films were grown on microscopic glass substrates kept at room temperature 
at distance of 12-15cm above the boat. Thin films were grown
at vacuum better than ${\rm 10^{-6}}$-${\rm10^{-7}}$Torr. Films grown by this 
method were found to be transparent or slightly translucent with a bluish or
bluish-green tinge depending on the films    
thickness. The films were removed from the chamber and immediately kept in 
a desiccator. All our studies were completed within five days of   
sample fabrication.

\section{Results and Discussion}

\subsection{Structural Studies}
Structural studies of the films were done using X-ray diffractometer 
(Philips PW 3020). The as grown films of CsCl were found to be polycrystalline 
in nature without exception. Characterizing peaks were found between
${\rm 2\theta=20-50^o}$ (fig~1). The peak positions match those listed in ASTM Card 
No.05-0607 confirming the chemical composition of the thin films to be that of 
CsCl. Cesium
Chloride exists in simple cubic state with a chlorine atom at the
center of the cube and one cesium atom at every corner of the unit
cell.\cite{kittel} 
The first peak is located around ${\rm 21.5^o}$ is Miller indexed 
as (100). Hence, the inter-planar spacing or the $'d'$ value of this peak is 
the lattice constant of unit cell itself. 
\begin{figure}[h]
\begin{center}
\epsfig{file=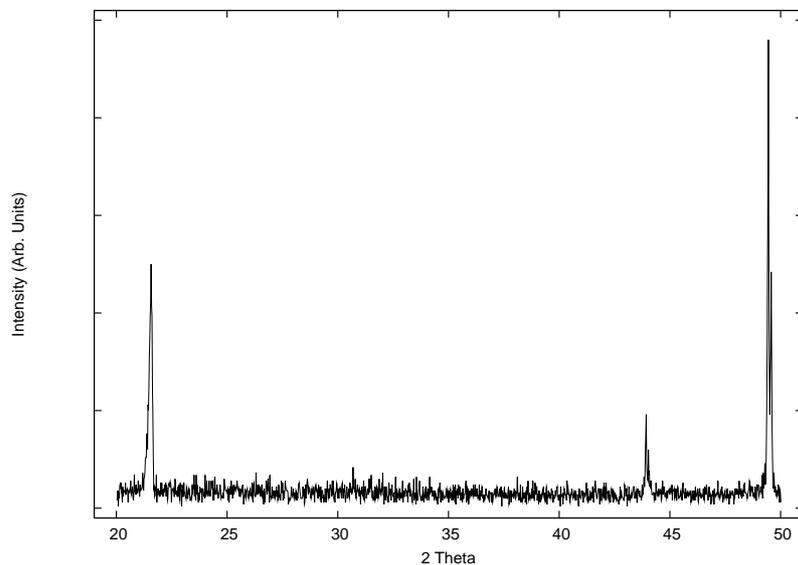,width=3.00in, angle=-90}
\vskip -.3cm
\caption{\sl X-Ray diffraction pattern of a CsCl thin film 
studied.}
\vskip -1cm
\end{center}
\label{fig:xrd}
\vskip -0.5cm
\end{figure}

\subsection{Optical Studies in Visible Region}

\begin{figure}[h]
\begin{center}
\epsfig{file=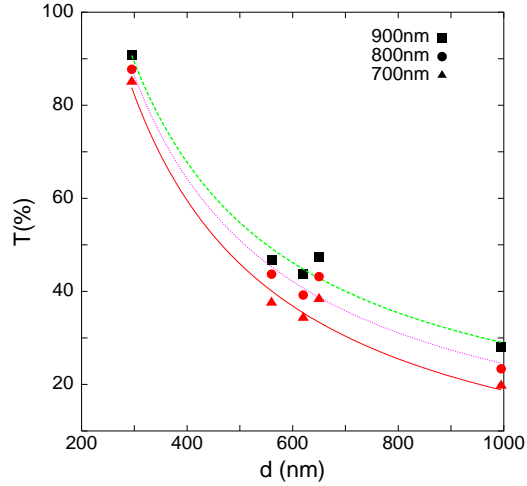,width=2.75in}
\vskip -.3cm
\caption{\sl The figure shows the
variation in transmission value with increasing film thickness at different
wavelengths. The nature of variation is same for all wavelengths with the best 
fit taking the form ${d=A+{B \over d}}$).}
\vskip -1cm
\end{center}
\label{fig:7}
\vskip -0.5cm
\end{figure}
We now detail the results of our optical studies
carried out using the UV-visible spectroscope (Shimadzu's UV 2501-PC). 
The refractive index of the thin films are usually calculated by employing 
Swanepoel's method.\cite{r220} This involves drawing two 
envelopes connecting the maxims and the minima are seen in the fringes of the 
transmission spectra. The interferences fringes seen in the transmission 
spectra is the result of multiple reflection from air-film and
film-substrate interface. However, the transmission spectra of CsCl films 
without exception did not show multiple and sharp fringes. This 
inevitably made estimation of the refractive index difficult. The lack of
multiple and sharp fringes in films may be due to the thickness of the thin 
film being very large or the refractive index of single layer films being of
the same order (nearly same values) as the substrate. Since the first 
possibility of all our films being too thick is ruled out, it can safely be
assumed that the refractive index of CsCl films is similar to the refractive 
index of our glass substrate. The refractive index of single crystalline CsCl 
is reported to be 1.6.\cite{r32} Pinholes with water vapor trapped in it would 
reduce the refractive index in the case of thin films. This would result in 
the refractive index of the thin film to lie between 1.5 and 1.6, which is very
near to the refractive index of the glass (i.e. ${\sim 1.5}$). Since the 
distinguishably
of film and substrate diminishes the lack of interference fringes is explained. 
However, the data of transmission spectra data for various samples reveal a 
dependence on film thickness (fig~2).

Absorption spectra of CsCl shows a increasing absorption as one proceeds
from ${\rm \lambda=900nm}$ to 300nm. Such trend have been reported by Rabin
et al\cite{rab} and Avakian et al\cite{ava} in CsCl single crystals. On
these sloped backgrounds they reported the appearance of peaks which
represented color-centers created by excess of alkali atoms, which in turn 
generate halogen vacancies, or by halogen vacancies created by X-irradiation. 
These vacancies give rise to energy levels within the forbidden band explaining 
peaks in absorption spectra and a color to the usually transparent crystal. 
In the referred works\cite{rab, ava} color-centers were created by irradiating 
the single crystals with X-rays. 
In unirradiated samples, the peaks due to color-centers are too faint. 
Even for irradiated samples, the peaks may be
faint and usually become prominent only when the pre-irradiated absorption
spectra is used as the baseline and subtracted from the spectra.\cite{rab}
However, these works do not comment on the sloped absorption background
which also shows variation on X-ray irradiation. 

\begin{figure}[h]
\begin{center}
\epsfig{file=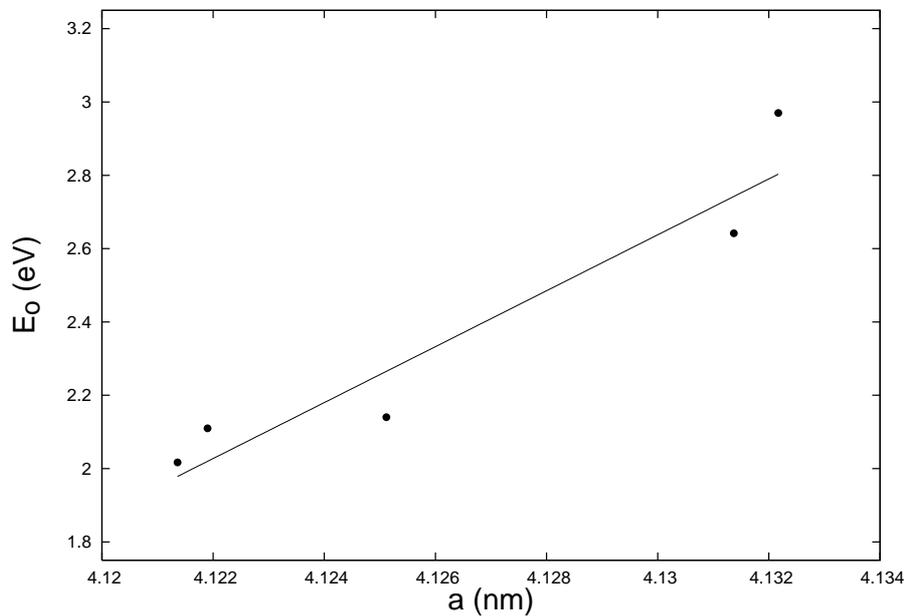,width=3.25in, angle=-90}
\vskip -.3cm
\caption{\sl The variation of the optical energy gap with lattice constants.}
\vskip -1cm
\end{center}
\label{fig:7al}
\vskip -0.5cm
\end{figure}

A slope in the absorption spectra is usually associated with multiple energy
levels coming together forming bands. Assuming, color centers to form narrow
bands within the forbidden band, we have used Tauc's\cite{r218} for
estimating band gap from the absorption spectra near band edges to determine
the difference in  energy level edge's (${\rm E_o}$) 
from the valence band to that of the
narrow bands coming in existence due to color-centers.  

We plot ${\rm E_o}$ as a function of lattice constants of the unit cells 
in fig~3. ${\rm E_o}$ associated 
with the color center film thickness. Films of thickness around 300nm has
energy of 3eV associated with it's color-center while the 1000nm thick film
has ${\rm E_o \sim 2eV}$, explaining the bluish to bluish-green tinge seen for 
the films with increasing film thickness. To analyze these results, we have 
modeled the point defects in our 
poly-crystalline films, with a cubic super-cell of 53 atoms with a Cl vacancy 
in its center. Because of the presence of Cs in the compound, we have included 
in our study the spin-orbit interaction. While this interaction splits up some 
energy bands, the main effects occur mainly in the intermediate valence band region 
(between -5.3 and -7.6 eV), 
which is far from the top of the valence band. Thus, it is not 
relevant in the energy region that determines the band gap. 

\begin{figure}[here]
\begin{center}
\epsfig{file=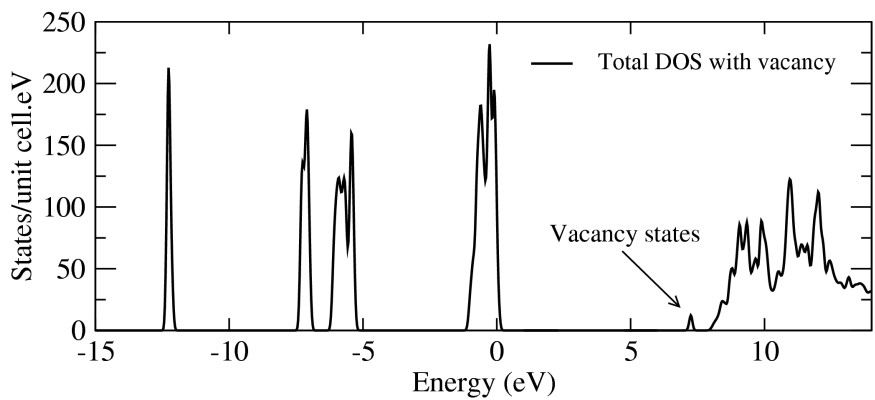,width=4.5in}
\vskip -.3cm
\caption{\sl Calculated density of states of the CsCl lattice with a Cl vacancy.}
\vskip -1cm
\end{center}
\label{fig:7}
\vskip -0.5cm
\end{figure}

Fig~4 shows the calculated density of states (DOS), where a flat band of 
localized states appears in the band gap, close to the bottom of the 
conduction band. Many of these vacancy states are filled with trapped electrons, 
turning the films active in the visible region. The minimum energy differences $E_0$ 
between these localized states to the first significant conduction band states 
are about 1.5 eV, and to the subsequent close first shoulders are about 1.6 eV, 
and 2.1 eV, in close agreement with the experimental F-centers 
energy of about 2.0 eV for bulk\cite{kittel} CsCl.
We have also calculated the energy position of the F-centers for 
different lattice parameter into the range at which our films were formed. 
The data points in fig~5 show theoretically calculated $E_0$ for 
different lattice parameters, which occur at about 
1.5 eV. In all cases, there are also close subsequent peaks up to about 2.0 eV 
(not shown in fig).  

\vskip .5cm
\begin{figure}[h]
\begin{center}
\epsfig{file=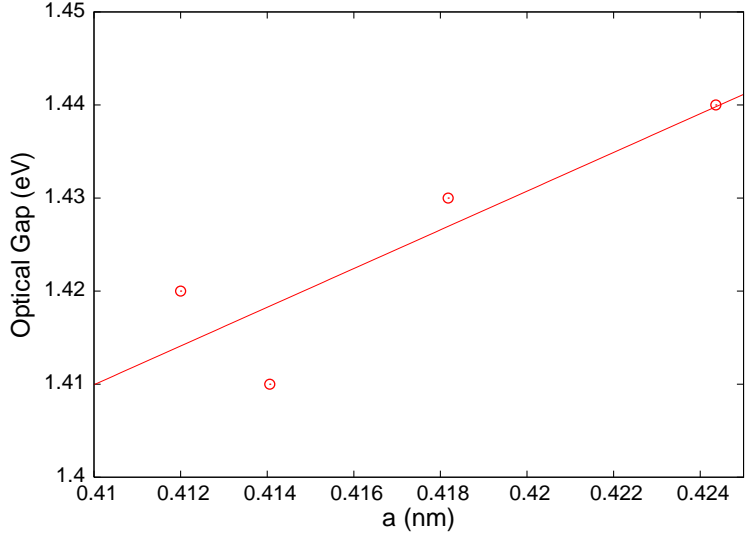,width=2.75in, angle=-90}
\caption{\sl Calculated optical energy gaps at different lattice parameters. 
}
\vskip -1cm
\end{center}
\label{fig:7}
\vskip -0.5cm
\end{figure}

\section{Conclusions}
Thin films of cesium chloride were fabricated on glass substrates held at room 
temperature. Due to CsCl's wide bandgap, these films should have been 
transparent, however, our film's had blue (very-light) to bluish green tinge which
suggests existence of color centers created by either excess Cs atoms or
vacancies due to absence of Cl atoms. These lead to variation in the unit
cell's lattice size, resulting in variation in the 
inter-atomic distances between neighboring Cs-Cl, Cs-Cs and Cl-Cl atoms. 
The band structure is also expected to change giving rise to altered optical 
properties of the films. We believe our theoretical and experimental results
highlight these material behavior. 

\section*{Acknowledgment}

The discussions with Dr. Bob Angelnier is gratefully acknowledged. The help in
completing the spectroscopic and diffraction analysis by Mr. Dinesh Rishi (USIC), 
Mr. Padmakshan and Mr. Rohtash (Department of Geology, Delhi University) is also 
acknowledged.
L.M. and E.A.A. acknowledge financial support from the Consejo Nacional de 
Investigaciones Cient\'{\i}ficas y T\'ecnicas (CONICET), 
the Universidad Nacional de Entre R\'{\i}os (UNER), and the Agencia Nacional 
de Promoci\'on Cient\'{\i}fica y Tecnol\'ogica (ANPCyT), Argentina.


\begin{thebibliography}{99}

\bibitem{rab} Herbert Rabin and James H. Schulman, Phys. Rev. {\bf 125},
1584 (1962).

\bibitem{ava} P. Avakian and A. Smakula, Phys. Rev. {\bf 120}, 2007 (1960).



\bibitem{WIEN2k}P. Blaha, K. Schwarz, J. Luitz, Viena University
of Technology (2001). (Improved and updated version of the WIEN code,
published by P. Blaha, K. Schwarz, P. Sorantin and S. B. Rickey, Comp.
Phys. Commun. 59, 399 (1990).

\bibitem{Perdew-etal92}J.P. Perdew, J.A. Chevary, S.H. Vosko, K.A. Jackson, 
M.R. Pederson, D.J. Sing, C. Fiolhais,  Phys. Rev. B \textbf{46} (1992-I) 6671.

\bibitem{Perdew-Wang92}J.P. Perdew, Y. Wang,  Phys. Rev. B \textbf{45} 
(1992-I) 13244.

\bibitem{PBE96}J.P. Perdew, K. Burke, M. Ernzerhof, Phys. Rev. Lett. 
77 (1996) 3865, and Phys. Rev. Lett. \textbf{78} (1997) 1396. 

\bibitem{Lanoo84}M. Lanoo, M. Schll\"uter, L.J. Sham, 
Phys. Rev. B \textbf{32} (1984) 3890. 

\bibitem{Wei-Zunger97}S. Wey, A. Zunger, Phys. Rev. B \textbf{55} (1977) 13605. 

\bibitem{Albanesi94}E.A. Albanesi, W.L. Lambrecht, B. Segall, Jour. of 
Vac. Science and Tech. B, \textbf{12} (1994) 2470. 

\bibitem{Laskowski-Christensen05}R. Laskowski, N.E. Christensen, G. Santi, and C.
Ambrosch-Draxl, Phys. Rev. B 72, 035204 (2005).

\bibitem{Nabi2000}Z. Nabi, B. Abbar, S. Mecabih, A. Khalfi, N. Amrane, 
Comput. Mat. Science \textbf{18} (2000) 127. 

\bibitem{kittel} C. Kittel, {\sl "Intoduction to Solid State Physics"}, 
John Wiley, NY(1953).

\bibitem{r2.7} B. D. Cullity, "{\sl Elements of X-Ray Diffraction}",
(Addison Wesley, New York, 1956).

\bibitem{r220} R. Swanepoel, J. Phys. E {\bf 16}, 1214 (1983).

\bibitem{r32} David R. Lide (ed), {\sl "CRC Handbook of Chemistry and
Physics"}, 
81st ed, CRC Press LLc Inc. USA (2000-2001).

\bibitem{r218} J. Tauc, {\sl "Amorphous and Liquid Semiconductors"}, J. Tauc Ed.
(Plenum, London, 1974).


\end{thebibliography}
\end{document}